\documentstyle[12pt]{article}

\def\be{\begin{equation}}
\def\ee{\end{equation}}
\def\ba{\begin{array}{c}}
\def\ea{\end{array}}

\begin{document}

\titlepage

 \begin{center}{\Large \bf
Generalized Rayleigh-Schr\"{o}dinger perturbation theory as a
method of linearization of the so called quasi-exactly solvable
models
  }\end{center}

\vspace{5mm}

 \begin{center}
Miloslav Znojil$^\dagger$\footnote{e-mail: znojil@ujf.cas.cz }
\vspace{3mm}

$\dagger$
 \'{U}stav jadern\'e fyziky AV \v{C}R,
250 68 \v{R}e\v{z}, Czech Republic

\end{center}

\vspace{5mm}

\section*{Abstract}

Sextic oscillator in $D$ dimensions is considered as a typical
quasi-exactly solvable (QES) model. Usually, its QES $N-$plets of
bound states have to be computed using the nonlinear and coupled
Magyari's algebraic equations. We propose and describe an
alternative linear method which is $N-$independent and works with
power series in $1/\sqrt{D}$. Main merit: simultaneous exact
solvability (for all the $N$ QES states) in the first two leading
orders (the degeneracy is completely removed, the unperturbed
spectrum is equidistant). An additional merit: All the
perturbation corrections are given by explicit matrix formulae in
integer arithmetics (there are no rounding errors).

 \vspace{9mm}

\noindent
 PACS 03.65.Ge


 \begin{center}
\end{center}

\newpage

\section{Introduction}

Sextic Hamiltonian in $D$ dimensions
 \[
 H = -\triangle + a\,|\vec{r}|^2
 + b\,|\vec{r}|^4 + c\,|\vec{r}|^6, \ \ \ \ \ \ \ \ \ a = a(N)
 \label{sextic}
 \]
enters many phenomenological and methodical considerations as a
``next-to-solvable" model \cite{Znojil:Classif}. In fact, among
all the real polynomial interactions, only the {\em harmonic} {\em
and sextic} models can generate an arbitrary $N-$plet of bound
state wavefunctions in an elementary form. All the similar models
are often called quasi-exactly solvable (QES,
cf.~\cite{Znojil:Ushveridze}).

Unfortunately, the close parallel between the sextic and harmonic
oscillator is not too robust and breaks down in practical
applications \cite{Znojil:Cannata}. For example, the
Rayleigh-Schr\"{o}dinger unperturbed propagator ceases to be
diagonal in the sextic case \cite{Znojil:prop}. Moreover, the key
weakness of {\em any} QES model lies in the nonlinearity of its
secular equation which has the polynomial form of degree $N$
\cite{Znojil:Singh}. Non-numerical determination of the sextic
energies is only feasible at $N \leq 4$. Otherwise, in a sharp
contrast to harmonic case, the values of energies $E_n$ are only
available up to some rounding errors.

In order to refresh the parallels we shall describe a new approach
to the sextic QES bound state problem. It is based on some
surprising results of the symbolic manipulation experiments. They
were performed in MAPLE using the technique of Groebner bases. We
revealed that the QES energies become equidistant and proportional
to integers in the limit of the large spatial dimensions $D \to
\infty$. This feature is presented in Sections~2~and~3.

In the second step of our analysis one discovers that the
systematic evaluation of the Rayleigh-Schr\"{o}dinger corrections
proves feasible in closed form. In spite of the non-diagonality of
propagators, a merely slightly modified form of construction can
be used. It gives the energy formula
  \[
  E(\lambda)=E^{(0)} + \lambda\,E^{(1)} +
 \lambda^2\,E^{(2)} +\ldots + \lambda^K\,E^{(K)} + {\cal
 O}(\lambda^{K+1}), \ \ \ \ \ \ \lambda = 1/\sqrt{D}.
 \]
Its coefficients $E^{(k)}$ are obtainable {\em without any
rounding errors} (cf. Sections~4 and~5 below).

\section{An unusual solvable limit: Large dimensions $D$}

All the sextic oscillator states are determined by the radial
Schr\"{o}dinger equation
 \begin{equation}
\left[-\,\frac{d^2}{dr^2} + \frac{\ell(\ell+1)}{r^2}+
 a\,r^2+b\,r^4 + c\,r^6\right]\,
\psi(r) = E\, \psi(r).
 \label{SExt}
 \end{equation}
It contains the dimension $D$ and the angular momenta $k = 0, 1,
\ldots$ in $\ell = k+(D-3)/2$. The elementary ansatz
 \begin{equation}
\psi(r) = \sum_{n=0}^\infty\, h_n \,r^{2n+\ell+1}\,{\rm
exp}\left(-\frac{1}{2}{\beta\,r^2 } - \frac{1}{4}{\gamma\,r^4
}\right), \ \ \ \ \ \  c=\gamma^2 > 0, \ b = 2 \beta \gamma
> 0
 \label{ana}
 \end{equation}
converts this ordinary differential equation into the linear
algebraic system characterized by the tridiagonal Hamiltonian
matrix,
\begin{equation}
Q^{[N]}\,\vec{h}=E\,\vec{h} , \ \ \ \ \ \ \ \   Q^{[N]}=
 \left(
  \begin{array}{ccccc}
 B_{0} & C_0&  & &  \\
A_1&B_{1} & C_1&    & \\ &\ddots&\ddots&\ddots&\\
&&A_{N-2}&B_{N-2}&C_{N-2}\\ &&&A_{N-1}&B_{N-1}
 \end{array}
 \right )
 \label{tridSE}
 \end{equation}
where the dimension is to be infinite, $N \to \infty$, and the
matrix elements are elementary,
\begin{equation}
\begin{array}{c}
 A_n = \gamma\,(4n+2\ell+1) + a -
\beta^2, \ \ \ \ \ \ \ B_n = B_n(E)=\beta\,(4n+2\ell +{3}),\\
 C_n = -2(n+1)\,(2n+2\ell+3),
  \ \ \ \ n = 0, 1, \ldots \ .
  \end{array}
  \label{elem2}
   \end{equation}
The (quasi-)variational limit $N \to \infty$ gives the numerically
correct spectrum \cite{Znojil:Hautot}. For the sake of simplicity,
let us now constrain our attention to the simplified model of
Singh et al \cite{Znojil:Singh} characterized by the QES condition
inposed upon the quadratic coupling $a = a(N)$,
 \[
  a(N)=
 \frac{1}{4\gamma^2}\,b^2-\gamma\,(4N+2\ell+1).
 \label{cond}
 \]
In this way one achieves the {\em rigorous} termination of the
wavefunctions,
 \begin{equation}
h_N=h_{N+1} =h_{N+2}=\ldots = 0.
 \label{trim}
 \end{equation}
The latter assumption merely changes the lower diagonal in eqs.
(\ref{tridSE}) and (\ref{elem2}) to the shorter formula
$A_n=4\gamma\,(n-N)$. Exact energies become available only at the
first few integers $N \leq 4$. Beyond $N=4$, QES solutions remain
numerical. Moreover, the intrinsic asymmetry of our Hamiltonian
(\ref{tridSE}) causes a loss of precision which grows quickly with
the degree~$N$~\cite{Znojil:Hautot}.

In such a setting we have noticed, purely empirically, that the
solutions are getting simpler when the spatial dimensions grow, $D
\gg 1$. In the leading-order approximation, the corresponding
matrix Schr\"{o}dinger equation becomes diagonally dominated,
\begin{equation}
 \label{tr}
 \left( \begin{array}{ccccc}
E-\beta D & 2D & & &  \\
 4(N-1)\gamma & E-\beta D& 4 D &  &  \\
 &\ddots&\ddots&\ddots&\\
&& 6\gamma & E-\beta D& 2 ({N-1})D   \\ &&& 4\gamma & E-\beta D
\end{array} \right)
 \left( \begin{array}{c}
 {h}_0\\
 {h}_1\\
\vdots \\
 {h}_{N-2}\\
 {h}_{N-1}
\end{array} \right )
= 0.
 \end{equation}
This enables us to evaluate the fully degenerate dominant
eigenvalue,
 \begin{equation}
E = \beta D - 2 \sqrt{2\gamma D}\,z
 \label{giving}
 \end{equation}
where $z$ is a constant.

\section{The removal of degeneracy in sub-dominant approximation}

Once we switch to the new energy variable $z$, we may pre-multiply
eq. (\ref{tr}) by a diagonal and regular matrix with elements
$\rho^{j}$ where $\rho =\sqrt{D/(2\gamma)}$. This leads to the
new, non-diagonal matrix Schr\"{o}dinger equation. It determines
the leading-order components of the renormalized Taylor
coefficients ${p}_j=[D/(2\gamma)]^{j/2}h_j$  and has the following
transparent form,
\begin{equation}
 \label{trap}
 \left( \begin{array}{ccccc}
0 & 1 & & &  \\
 (N-1) & 0& 2 &  &  \\
 &\ddots&\ddots&\ddots&\\
&& 2 & 0&  ({N-1})   \\ &&& 1 & 0
\end{array} \right)
 \left( \begin{array}{c}
 {p}_0\\
 {p}_1\\
\vdots \\
 {p}_{N-2}\\
 {p}_{N-1}
\end{array} \right)
=
z \cdot
 \left( \begin{array}{c}
 {p}_0\\
 {p}_1\\
\vdots \\
 {p}_{N-2}\\
 {p}_{N-1}
\end{array} \right ).
 \end{equation}
In spite of the manifest asymmetry of this equation, all its
eigenvalues remain strictly real. We computed these eigenvalues by
symbolic manipulations in integer arithmetics and discovered that
the underlying nonlinear secular equation is solvable exactly and
completely. The $N-$plets of its energy roots proved
nondegenerate, equidistant and extremely elementary,
 \begin{equation}
\left ( z_1, z_2, z_3, \ldots, z_{N-1}, z_N \right ) = \left (
-N+1, -N+3, -N+5,  \ldots, N-3, N-1 \right ).
 \label{mainr}
 \end{equation}
This result is valid {\em at an arbitrary finite matrix size $N$}.

It is quite elementary to verify that also the the respective left
and right eigenvectors remain real. Up to their norm, all of them
can be represented in terms of integers. Their components may be
arranged in the rows and columns of certain square matrices,
 \[
 P(0) = 1, \ \ \ \ \ \ \ P(1)=
\frac{1}{\sqrt{2}} \,
 \left( \begin{array}{rr}
1&1\\ 1&-1
\end{array} \right),
 \]  \[  P(2)= \frac{1}{\sqrt{4}} \,
 \left( \begin{array}{rrr}
1&1&1\\ 2&0&-2\\ 1&-1&1
\end{array} \right),
 \ \ \ \ \ \ \
P(3)= \frac{1}{\sqrt{8}} \,
 \left( \begin{array}{rrrr}
1&1&1&1\\ 3&1&-1&-3\\ 3&-1&-1&3\\ 1&-1&1&-1
\end{array} \right),
 \]
 \[  P(4)= \frac{1}{\sqrt{16}} \,
 \left( \begin{array}{rrrrr}
1&1&1&1&1\\ 4&2&0&-2&-4\\ 6&0&-2&0&6\\ 4&-2&0&2&-4\\ 1&-1&1&-1&1
\end{array} \right)
 \]
etc. These matrices $P=P(N-1)$ are all asymmetric but idempotent,
$P^2=I$.

We may summarize that in the limit $D \to \infty$, the QES sextic
model may be factorized easily. After a suitable normalization,
all the components of the eigenvectors are integers.

\section{An adapted Rayleigh-Schr\"{o}dinger perturbation recipe}

At the finite values of $D$ and {\em starting directly from the
second-order precision of preceding section}, the routine
perturbation theory becomes applicable since the unperturbed
Hamiltonian remains diagonal and  all its spectrum is safely
non-degenerate.


At any $D \gg 0$ the Schr\"{o}dinger equation (\ref{tridSE}) is an
eigenvalue problem with the perturbed Hamiltonian of the two-term
form,
 \[
H(\lambda)=H^{(0)} + \lambda\,H^{(1)} + \lambda^2H^{(2)}, \ \ \ \
\ \lambda = 1/\sqrt{D}
 .
 \]
Both the perturbations are one-diagonal matrices which depend on
the value of the angular momentum $k$,
 \[
\left ( H^{(1)} \right )_{nn}= \frac{\beta}{\sqrt{2\gamma}}
(2n+k), \ \ \ \ \ \ n = 0, 1, \ldots, N-1, \]  \[  \left ( H^{(2)}
\right )_{nn+1}= -(n+1)(2n+2k), \ \ \ \ \ \ n = 0, 1, \ldots, N-2.
 \]
We may re-write our Schr\"{o}dinger equation (\ref{tridSE}) in the
textbook perturbation-series representation at any $N$,
 \begin{equation}
 \begin{array}{c}
\left ( H^{(0)} + \lambda\,H^{(1)} + \lambda^2H^{(2)} \right )
\cdot \left ( \psi^{(0)} + \lambda\,\psi^{(1)} + \ldots +
 \lambda^K \psi^{(K)}
+{\cal O}(\lambda^{K+1}) \right )\\
=
\left ( \psi^{(0)} + \ldots +
 \lambda^K \psi^{(K)}
+{\cal O}(\lambda^{K+1}) \right ) \cdot \left ( \varepsilon^{(0)}
 +
\ldots +
 \lambda^K \varepsilon^{(K)}
+{\cal O}(\lambda^{K+1}) \right ). \end{array} \label{pertu}
 \end{equation}
Let us again concatenate the (lower-case) zero-order vectors $
\vec{p}=\vec{p}^{(0)} \equiv \psi^{(0)} $ into an $N$ by $N$
matrix $P=P^{(0)}$, with all the eigenvalues arranged also in a
diagonal matrix $\varepsilon^{(0)}$. In this way the zero-order
equation $H^{(0)} \psi^{(0)} =\psi^{(0)} \varepsilon^{(0)} $ is
satisfied identically. Indeed, in our compactified notation, it
reads $ P\varepsilon^{(0)}PP = P\varepsilon^{(0)}$ and we know
that $P^2=I$.

With the factorized $H^{(0)} = P\varepsilon^{(0)}P$, we shall use
the same convention in all orders and concatenate the vectors
$\vec{\psi}^{(k)}_j, \,j=1, 2, \ldots, N$ in the square matrix
$\Psi^{(k)}$. In the first order of perturbation analysis this
replaces the ${\cal O}(\lambda)$ part of eq. (\ref{pertu}) by the
matrix relation
 \begin{equation}
\varepsilon^{(1)} + P\,\Psi^{(1)} \varepsilon^{(0)}
-
\varepsilon^{(0)} P\, \Psi^{(1)}
=
P\,H^{(1)}\,P. \label{prvnirad}
 \end{equation}
In the second order we get
 \begin{equation}
\varepsilon^{(2)} + P\,\Psi^{(2)} \varepsilon^{(0)}
-
\varepsilon^{(0)} P\, \Psi^{(2)}
=
P\,H^{(2)}\,P + P\,H^{(1)}\,\Psi^{(1)}
-
P\,\Psi^{(1)} \varepsilon^{(1)} \label{druhyrad}
 \end{equation}
etc. The available expressions occur on the right-hand side of
these equations while the unknown quantities stand to the left.
All the higher-order formulae have the same structure.

We may summarize that the diagonal part of equations
(\ref{prvnirad}) or (\ref{druhyrad}) determines the energy
corrections $\varepsilon^{(1)}$ and $\varepsilon^{(2)}$,
respectively. Non-diagonal components of these matrix relations
are to be understood as a definition of the eigenvectors.

\section{Merits of the method: An $N=2$ illustration}

One has to move up to the higher-order level for the elimination
of the normalization ambiguities. This has been multiply clarified
in the literature on perturbation theory \cite{Znojil:cz}. Still,
we should emphasize a user-friendliness of this normalization
freedom within the framework of the present formalism. For
illustration, let us consider just the $s-$wave problem in the
$N=2$ case. Immediately, our first-order formulae give the two
energy corrections which are both equal to each other,
 \begin{equation}
 \varepsilon^{(1)}_{11}
=\varepsilon^{(1)}_{22}= \beta/\sqrt{2\gamma} .
 \label{next}
 \end{equation}
One discovers that the ${\cal O}(\lambda)$ level of precision
provides just an incomplete information about the norms of the
first-order wave functions. This is the well know normalization
freedom manifesting itself in the present setting. On the ${\cal
O}(\lambda)$ level of precision only two constraints $
\Psi^{(1)}_{11} -\Psi^{(1)}_{21} =-\beta/\sqrt{2\gamma}$ and $
\Psi^{(1)}_{12} +\Psi^{(1)}_{22} =\beta/\sqrt{2\gamma}$ are
imposed upon the wavefunctions. Their definition must be completed
in the subsequent order.

In any higher order computation, the use of the computerized
symbolic manipulations is strongly recommended. Their
implementation is trivial. The algorithm can be written in integer
mathematics and generates, therefore, the perturbation series
without any errors. This is our most important conclusion. One
generalizes immediately the above leading-order results
(\ref{giving}), (\ref{mainr}) and (\ref{next}) to the compact
energy series for our particular sextic $k=N-2=0$ illustration,
 \begin{equation}
E_{1,2} = \frac{\beta}{\lambda^2}
 \pm
\frac{2\sqrt{2\gamma}}{\lambda}
 +2\beta
 \pm
\frac{\beta^2}{\sqrt{2\gamma}}\,\lambda + 0\cdot \lambda^2
 \mp
\frac{\beta^4}{8\gamma\sqrt{2\gamma}}\,\lambda^3 + 0\cdot
\lambda^4
 + {\cal O}(\lambda^5).
 \label{givg}
 \end{equation}
One can observe the (complete) leading-order degeneracy of
section~2 as well as its immediate next-order removal
(\ref{mainr}) as discussed in section~3. It is also amusing to
notice the above, hand-evaluated and quite unexpected, degeneracy
of the subsequent ${\cal O}(1)$ correction.

One can notice the existence of certain identically vanishing
corrections here. In fact, their rigorous evaluation would not be
possible within the standard framework of perturbation theory
where the summations over the intermediate states must be computed
in finite precision. Only within the present formalism which is
able to work in integer arithmetics, the unusual feasibility of
proving the {\it precise cancellation} of the series of
corrections can be achieved. This is one of the less expected
though most important merits of our present methodical proposal
and construction.

\subsection*{Acknowledgements}

Work supported by the GA AS CR grant Nr. A 104 8004.

\end{document}